\documentstyle[12pt,epsfig]{article}
\begin{document}
\font\ninerm = cmr9
% \baselineskip 28pt plus .5pt minus .5pt
% \pageno=0\footline={\ifnum\pageno>0 \hss --\folio-- \hss \else\fi}

\def\footnoterule{\kern-3pt \hrule width \hsize \kern2.5pt}

\pagestyle{empty}

% \begin{flushright}
% astro-ph/0204077 \\
% August 2000
% \end{flushright}

\vskip 0.5 cm

%\vskip 1.5 cm
\begin{center}
{\large\bf Anomalous particle-production thresholds through
systematic and non-systematic quantum-gravity effects}
\end{center}
\vskip 1.5 cm
\begin{center}
{\bf Giovanni AMELINO-CAMELIA}$^a$, {\bf Y.Jack NG}$^b$
and {\bf Hendrik VAN DAM}$^b$\\
\end{center}
\begin{center}
{\it $^a$Dipart.~Fisica,
Univ.~Roma ``La Sapienza'',
P.le Moro 2, 00185 Roma, Italy}\\
{\it $^b$Institute of Field Physics,
Department of Physics and Astronomy,\\
University of North Carolina,
Chapel Hill, NC 27599-3255, USA}
\end{center}

\vspace{1cm}
\begin{center}
{\bf ABSTRACT}
\end{center}

{\leftskip=0.6in \rightskip=0.6in
A growing number of studies is being devoted to the identification
of plausible quantum properties of spacetime which might give
rise to observably large effects. The literature on this subject is
now relatively large, including studies in string theory,
loop quantum gravity and noncommutative geometry.
It is useful to divide the various proposals
into proposals involving a systematic quantum-gravity effect
(an effect that would shift the main/average prediction for
a given observable quantity) and proposals involving a
non-systematic quantum-gravity effect
(an effect that would introduce new fundamental uncertanties
in some observable quantity).
The case of quantum-gravity-induced particle-production-threshold
anomalies,
%an intensely
a much
studied example of potentially observable
quantum-gravity effect, is here used as an example to illustrate the
differences to be expected between systematic and non-systematic
effects.
}

%\pacs{{\tt$\backslash$\string pacs\{\}} }

\newpage

\baselineskip 12pt plus .5pt minus .5pt

\pagenumbering{arabic}
\pagestyle{plain}

\section{Introduction}

Recently, there has been increasing interest
in ``Quantum-Gravity Phenomenology"~\cite{polonpap,iucaapap},
the research programme that attempts to identify aspects
of quantum gravity that can be tested experimentally and
to identify
some of the corresponding experiments that can do significant
tests with presently-available technological capabilities.
Examples of possible experimental
searches have been considered in
Refs.~\cite{qgcpt,grbgac,billetal,gacgwi,nggwi,gacgwiLATEST,kifu,gactp,ngthresh,gacpion}
and (with an emphasis not directly related to quantum properties
of space-time) in
Refs.~\cite{polonpap,veneziano}.
Recent reviews of the status of this field can be found
in Refs.~\cite{rovhisto,iucaapap}.

Here we want to draw a sharper distinction between analyses that consider
``systematic" quantum-gravity effects and analyses that consider
``non-systematic" quantum-gravity effects.
The theoretical prediction that does not include the effects of
quantum gravity is characterized, with respect to a given
observable ${\hat X}$, by a prediction $X$ and, possibly, a fundamental (quantum
mechanical) uncertainty $\delta X$. The effects of quantum gravity
in general then should lead~\cite{iucaapap} to a new
prediction $X'$ and a new uncertainty $\delta X'$.
One would attribute to quantum gravity
the effects $(\Delta X)_{QG} \equiv X' - X$
and $(\delta X)_{QG} \equiv \delta X' - \delta X \geq 0$.
One can speak of purely systematic
quantum-gravity effect when $(\Delta X)_{QG} \neq 0$
and $(\delta X)_{QG} = 0$,
while the opposite case, $(\Delta X)_{QG} = 0$ and $(\delta X)_{QG} > 0$,
can be qualified as
purely non-systematic. One might expect both types of effect to be present
simultaneously, as it is emerging from certain developing
formalisms~\cite{gacdsr}, but in a given phenomenological programme
the focus may be put on one or the other case. For example, the type of
in-vacuo dispersion studies described in Refs.~\cite{grbgac,billetal} focuses
on a purely systematic effect, while the space-time-foam interferometric
studies described in Ref.~\cite{gacgwi,nggwi,gacgwiLATEST} focus on a
purely non-systematic effect.

We intend to characterize more carefully the differences
between these two types of quantum-gravity effects,
even though, depending on the specifics of
the physical context, the signatures to be sought experimentally
may or may not be sharply different.
Our analysis will mainly rely on the case study provided
by one of the possible quantum-gravity predictions which has recently
attracted most attention: the possibility of quantum-gravity
implications for the kinematical threshold for certain
particle-physics
%effects.
processes.
In that context, which we will review in the next section,
both types of effects can be meaningfully considered.
We will discuss their different manifestations,
and in particular we will argue that the characteristic
signatures of nonsystematic effects have a stronger
tendency to depend on some nontrivial features of the underlying
dynamical theory, whereas in some cases systematic effects
can be studied with the exclusive guidance of symmetry principles.
These points we will articulate in Section 3,
while Section 4 is devoted to some closing remarks.

\section{Quantum-gravity-induced threshold anomalies}

In two different regimes, UHECRs and multi-TeV photons, the
universe appears to be more transparent than expected.
UHECRs should interact with the Cosmic Microwave Background
Radiation (CMBR) and produce pions. TeV photons should interact
with the
% Far Infra Red Background (FIRB)
Far Infra Red Background (FIRB)
photons and produce electron-positron
pairs.  These interactions should make observations of UHECRs
with $E > 5 {\cdot} 10^{19}$eV (the GZK limit)\cite{GZK} or of
gamma-rays with $E > 10$TeV\cite{NGS} from distant sources
unlikely.  Still UHECRs above the GZK limit and
Mk501 photons with energies up to 24 TeV have been observed.

A CMBR photon and a UHE proton with $E >5 {\cdot} 10^{19}$eV
should satisfy the kinematic
requirements (threshold) for pion production.
UHE protons
should therefore loose energy, due to photopion production, and
should slow down until their energy is below the GZK energy. At
higher energies  the proton's mean free path decreases rapidly
and it is down to a few Mpc at $3 {\cdot} 10^{20}$eV. Yet more
than 15 CRs have been observed with nominal energies at or above
$10^{20} {\pm} 30\%$ eV\cite{AgaWat}. There are no astrophysical
sources capable of accelerating particles to such energies within
a few tens of Mpc from us. Furthermore, if the CRs are produced
homogeneously in space and time, we would expect a break in the CR
spectrum around the GZK threshold, which is not seen.

HEGRA has detected high-energy photons with a spectrum ranging up
to 24 TeV\cite{Aharonian99} from Markarian 501 (Mk501), a BL Lac
object at a redshift of 0.034 ($\sim 157$ Mpc). This observation
might indicate a second paradox of a similar nature. A high energy
photon with energy $E$ can interact with a FIRB photon
with wavelength $\lambda \sim 30\mu m  (E/10TeV)$ and produce
an electron-positron pair. The mean free path of TeV photons depends
on the spectrum of the FIRB.
Recent data from DIRBE\cite{Wright00,Finkbeiner,Hauser98}
and from ISOCOM~\cite{Biviano99}
suggest that the mean free path for 20TeV
photons should be much shorter than the one of 10TeV photons.
However, no apparent break is seen in the spectrum of Mk501
in the region $10$-$20$TeV.

The UHECR paradox is well established. Numerous theoretical
models, mostly requiring new physics, have been proposed for its
resolution (see Ref.\cite{Olinto} for a recent review). With
much less data, and with some uncertainty on the FIRB,
the Mk501 TeV-photon paradox is less established\footnote{For an analysis
of the Mk501 experimental situation that does not advocate
threshold anomalies see, {\it e.g.}, Ref.~\cite{steckernoparadox}.}.
However, if indeed this must be considered as a paradox,
there are no models for its resolution,
apart from the possibility that the FIRB estimates
are too large. Planned experiments will soon
provide us better data on both issues.
At present it appears reasonable to explore the possibility
that both paradoxes are real.

One natural interpretation of these paradoxes
relies on quantum gravity effects~\cite{kifu,gactp,ngthresh}:
both classes of puzzling
observations, cosmic rays and Mk501 photons,
at the theoretical level involve the evaluation of
a particle production threshold.
The relevant threshold conditions are nothing but a
statement about the symmetries of classical Minkowski spacetime.
It is therefore conceivable
that one might explain the puzzling observations in terms of a
quantum-gravity-induced ''threshold anomaly"~\cite{gactp}, i.e.,
a different prediction for thresholds which might result
from quantum properties of spacetime.
In both paradoxes, low-energy photons
interact with high energy particles. The relevant reactions should take
place at a kinematic threshold. In both cases, the center-of-mass threshold
energies are rather modest and the physical processes involved are well
tested and understood.
%In spite of these similarities,  so far, there is no
%model that explains both paradoxes within a single theoretical scheme
%(unless the model accommodates an irritatingly large number of parameters).
These observations appears to encourage the hypothesis that
quantum-gravity-induced deviations from ordinary Lorentz invariance might
be responsible for both paradoxes\footnote{It is of course also possible to
consider deviations from Lorentz invariance that do not have
quantum-gravity origin\cite{Gonzalez}, but, as discussed in
Refs.\cite{gactp}, the idea of a quantum-gravity origin, besides being
conceptually appealing, leads to a natural estimate for the magnitude of
the effects, and this estimate appears to fit well the observations (while
non-quantum-gravity approaches host a large number of parameters which have
to be
freely adjusted to obtain the needed magnitude of departures from
Lorentz-invariance).}.

Most studies have considered the possibility that quantum-gravity effects
might shift the threshold systematically (in the language adopted in the
introduction, this would be a ''systematic" quantum-gravity effect). In
order to illustrate this systematic mechanism of threshold anomalies, let
us consider, for example, the case in which
quantum-gravity effects are assumed
to induce a deformation of the
dispersion relations of the type
\begin{equation}
E^2 \simeq  \vec{p}^2 + m^2
+ \eta \vec{p}^2 \left({E \over
E_{p}}\right)^\alpha
~,
\label{dispone}
\end{equation}
where $E$ and $\vec{p}$ are
the energy and the (3-component) momentum of the particle, $E_{p}$
is the Planck energy scale ($E_{p} \sim 10^{16}$TeV), $\alpha$
and $\eta$ are free parameters characterizing the deviation
from ordinary Lorentz invariance.\footnote{Space-time structures
leading to (\ref{dispone})
do not necessarily reflect a loss of symmetry with respect to
ordinary Minkowski space-time.
For example, with $\alpha = 1$, (\ref{dispone}) can
characterize certain quantum spacetimes as an observer-independent
property~\cite{gacdsr}, and in those spacetimes one still has a
(modified) concept
of Lorentz invariance which amounts to as many symmetries as in the
ordinary classical-spacetime case.
In this Section we are, however, effectively considering a broken-symmetry
case; in fact, we assume ordinary conservation of
energy-momentum, while the case in which
a dispersion relation (\ref{dispone}) does not lead to symmetry loss
involves~\cite{gacdsr} a corresponding deformation of
energy-momentum conservation.}

Experimental tests of the predictions of (\ref{dispone})
were first proposed in Ref.~\cite{grbgac},
for the case $\alpha=1$,
and in Ref.~\cite{polonpap}, for generic $\alpha$.

Let us consider the implications of (\ref{dispone})
for the kinematics of the process of
electron-positron pair production, which is relevant for the Mk501 paradox.
Combining (\ref{dispone}) with ordinary energy-momentum conservation
rules, one can establish that, for the case $\alpha = 1$,
at threshold the energy $E$ of the hard Mk501
photon and the energy $\epsilon$ of the soft background photon satisfy
the relation $E \simeq m_e^2/ \epsilon - \eta m_e^6/(8 E_p \epsilon^4)$.
For negative $\eta$ (and $|\eta| \sim 1$), the
correction $- \eta m_e^6/(8 E_p \epsilon^4)$ is indeed sufficient to push the
threshold energy upwards by a few TeV, consistently with observations.
As shown in Refs.\cite{gactp} (and references therein), an analogous result
holds for the photopion threshold, which is relevant for the cosmic-ray
paradox.

This type of analysis shows
that the two paradoxes might indeed
be a manifestation of a quantum (Planck-length related)
property of spacetime.

We have so far assumed that the quantum-gravity effect responsible for the
puzzling observations in cosmic-ray and Mk-501 physics is of the systematic
type, but in Ref.~\cite{ngthresh} some non-systematic
quantum-gravity effects were considered in relation with these same
observations. Our main objective here is to compare the
non-systematic effects with the already
understood systematic effects.

\section{Non-systematic effects and threshold anomalies}

In the previous Section, we have reviewed the use of a systematic
quantum-gravity effect, a systematically deformed dispersion relation,
for the solution of threshold anomalies.
One can easily construct an alternative scheme in which the systematic
effect responsible for the anomalies appears in energy-momentum
conservation (or both in the dispersion relation and in energy-momentum
conservation), but we shall not be concerned with these alternatives,
since our focus is on distinguishing
between systematic and non-systematic effects.

Accordingly, in this Section, we consider the case of a non-systematic
quantum-gravity-induced deformation of the dispersion relation,
more specifically, the case in which the classical relation $E^2 = p^2 + m^2$
still holds on average, but in a given physical realization $E$ will be related
to $p^2$ in a way that can be slightly different from
the average relation: $E^2 = p^2 + m^2 + \Delta$,
with $- \eta \vec{p}^2 ( E / E_{p})^\alpha
< \Delta < \eta \vec{p}^2 (E / E_{p})^\alpha$.

As we shall see, the implications of these non-systematic
effects may depend heavily on
the mechanism by which the uncertainty (the non-systematic effect)
is attributed in a given physical process.
In preparation for that discussion it is useful to consider
a simplified problem, which captures some aspects of the
threshold-anomaly problem here of interest.

\subsection{A useful analogy}

The observation of cosmic-rays or photons above threshold
is puzzling because they come from far away
and they should have disappeared (or "degraded", in
the case of cosmic rays) along the way, by encountering
appropriate "sinks" (the soft background photons).

For the purpose of the analysis we are reporting here,
it is useful to consider a much simplified analogous problem.
Consider a setup in which wheels,
of diameter $D_w$,
are sent from point $A$ to point $B$.
On the path from $A$ to $B$, there are
some holes of diameter $D_h$.
If $D_w < D_h$ the wheels will disappear in the first
hole encountered and will not reach point $B$.
If $D_w > D_h$ the wheels will safely reach point $B$.

One can speak of a threshold anomaly for the case in which
the classical predictions for $D_w$ and $D_h$ are such
that $D_w < D_h$, but some wheels still reach
point $B$ from $A$. We are assuming that the number of wheels
sent from $A$ toward $B$ per unit time is not known.
In the case of the "classical" prediction $D_w < D_h$
the ignorance about this frequency of emission is not
significant: in the case $D_w < D_h$ no wheels should reach $B$.
If some wheels are observed at $B$ something must be wrong
with the "classical theory".

Such an anomaly could be explained as a systematic "non-classical"
effect: one could argue that either the correct (non-classical)
value of the wheel diameter $D_w'$ is such that $D_w' > D_h > D_w$
or that the correct value of the hole size $D_h'$ is
such that $D_h' < D_w < D_h$
(in other words one should assume that the classical estimate,
$D_w$, $D_h$, is systematically incorrect, and that the correct
values of wheel diameter and the whole size are some $D_w'$, $D_h'$
such that $D_h' < D_w'$).

One can also advocate non-systematic effects. Assume for example
that, while the classical estimate of the hole size $D_h$ is correct,
the correct non-classical description of the mechanism
that sends wheels from $A$ to $B$ is such that not all wheels have the
same diameter $D_w$, but any given wheel has a diameter $D_w'$
with $D_w - \Delta < D_w' < D_w + \Delta$,
according to some probability law.
Then it would not be surprising to observe some wheels arriving at $B$,
since some of the wheels do have a diameter $D_w'$ such that $D_w' > D_h$.
This would provide an explanation of the threshold anomaly.\footnote{This
is analogous to the non-systematic effect advocated in Ref.~\cite{ngthresh}.}

On the other hand, if the classical
estimate of $D_w$ is correct but a non-classical uncertainty
affects $D_h$, one does not necessarily obtain an explanation
of the threshold anomaly. The holes found on the way between
point $A$ and point $B$ may not all have the same size $D_h$,
but instead the i-th hole has size $D_h^i$ with value
somewhere in the range $D_h - \Delta < D_h^i < D_h + \Delta$,
according to some probability distribution.
In this case, if only one (or a few) hole is encountered
on the way from $A$ to $B$, there will be a significant portion of
the wheels that reachs $B$, providing an explanation for
the threshold anomaly.
But if the number of holes found on the way from $A$
to $B$ is large, then there will only be a negligibly small
probability for a wheel to reach point $B$; in fact,
out of many holes there will easily be at least one
with size $D_h^i$ such that $D_w < D_h^i$.
A non-systematic effect, even one that at first sight might appear
well suited for providing an explanation for the threshold
anomaly, is not necessarily sufficient to solve
the threshold anomaly.

In closing this Subsection, let us consider a scenario
in which both systematic
and non-systematic effects coexist. Let us assume that the classical
estimate $D_w$ of the size of the wheels is incorrect and must be replaced
systematically with the estimate $D_w'$, and that $D_w'>D_h$. Moreover, let
us assume that the size of the holes is $D_h$ only on average,
but the i-th hole has size $D_h^i$ with value somewhere
in the range $D_h - \Delta < D_h^i < D_h + \Delta$,
according to some probability distribution.
In such a situation it is crucial to establish whether
$D_w'$ is greater than $D_h + \Delta$, because,
in spite of the fluctuations in the sizes of the holes, this would mean that
the wheels can get safely from $A$ to $B$.
But, if $D_w' < D_h + \Delta$, and the number of holes is large,
it is then likely that the wheel will encounter on the way from $A$
to $B$ at least one hole large enough for the wheeel to disappear
into it. This scenario is interesting because it shows that,
for theories with both systematic and nonsystematic effects, one
should really do a careful analysis: it is not sufficient to
analyze the systematic effects. We have seen, via this analogy,
that non-systematic effects can in some situations give
rise to a systematic result (in our analogy, if the number of
holes is large the wheel will disappear into one of the holes,
nearly systematically). And non-systematic effects can counter-act
the action of systematic effects (in our analogy, the systematic
effect in the wheel diameter, which would have explained the
threshold anomaly, is offset by the non-systematic effect
on the sizes of the holes).

\subsection{Threshold anomalies and non-systematic
quantum-gravity effects}

In Section~2 we have reviewed the most common use
of quantum-gravity ideas in relation with the threshold
anomalies: a systematic quantum-gravity effect
on the dispersion relation, which affects the threshold
prediction.
In Ref.~\cite{ngthresh} another type of quantum-gravity effects
was preliminarily considered in relation with the threshold anomalies:
the possibility of a new uncertainty principle between energy
and momentum.

Motivation for these new uncertainty principles can be found
in Ref.~\cite{padma,ngthresh}. For the purposes
of the present study it is significant that a new uncertainty
principle introduces new non-systematic effects
in quantum-gravity phenomenology.
The analogy drawn in the previous Subsection will help
us quickly analyze the possible implications of such
non-systematic effects for threshold anomalies.

For simplicity, let us focus on some non-systematic effects that are
the natural counterpart for the systematic effects
described by Eq.~(\ref{dispone}) for the case $\alpha = 1$.
Whereas, according to Eq.~(\ref{dispone}), the relation between
energy and momentum at high energies is systematically
deformed to
\begin{equation}
E \simeq  |\vec{p}| + {m^2 \over 2 |\vec{p}|}
+ {\eta \over 2} {\vec{p}^2  \over E_{p}}
~,
\label{dispsyst}
\end{equation}
in an uncertainty-relation-based scheme for non-systematic quantum-gravity
effects one can assume that on average the classical-spacetime relation
between energy and momentum is still satisfied, but for a given
particle with large
momentum $\vec{p}$, energy would be somewhere in the range
\begin{equation}
|\vec{p}| + {m^2 \over 2 |\vec{p}|}
- { |\eta| \over 2} {\vec{p}^2  \over E_{p}}
\le E \le
|\vec{p}| + {m^2 \over 2 |\vec{p}|}
+ {|\eta| \over 2} {\vec{p}^2  \over E_{p}}
~,
\label{dispnons}
\end{equation}
with some (possibly gaussian) probability distribution.
The given quantum-gravity theory will host a fundamental
value of $\eta$ (reasonably conjectured to be of
order $1$~\cite{grbgac,gacdsr}), but each given particle will
satisfy a dispersion relation of the type
\begin{equation}
E \simeq  |\vec{p}| + {m^2 \over 2 |\vec{p}|}
+ {{\tilde {\eta}} \over 2} {\vec{p}^2  \over E_{p}}
~,
\label{disptilde}
\end{equation}
with $-|\eta| \le {\tilde {\eta}} \le |\eta|$.

Let us again focus on the example of electro-positron pair production
in a head-on photon-photon collision, again assuming that (as relevant for
the analysis of Mk501 observations) one of the photons is very hard
while the other one is very soft.
To leading order, the value of ${\tilde {\eta}}$
for the given soft photon is not important, only the energy $\epsilon$
of the soft photon is significant. The soft photon
can, in leading order, be treated as satisfying a classical
dispersion relation.
In a quantum-gravity theory predicting
such non-systematic effects, the hard photon is mainly characterized
by its energy $E$
and its value of ${\tilde {\eta}}$. In order to establish whether
a collision between two such photons can produce an electron-positron
pair, one should establish whether, for some admissable values
of ${\tilde {\eta}}_+$ and ${\tilde {\eta}}_-$ (the values
of ${\tilde {\eta}}$ pertaining to the positron and the electron
respectively), the conditions for energy-momentum conservation
can be satisfied.
The process will be allowed if
\begin{equation}
E \ge  {m^2 \over \epsilon}
- {{\tilde {\eta}} \over 4} {E^3  \over \epsilon E_{p}}
+ {{\tilde {\eta}}_+ + {\tilde {\eta}}_- \over 16} {E^3  \over \epsilon E_{p}}
~.
\label{thr1prelim}
\end{equation}
Since ${\tilde {\eta}}$, ${\tilde {\eta}}_+$ and ${\tilde {\eta}}_-$
are bound to the range between $-|\eta|$ and $|\eta|$,
the process is only allowed, independently of the
value of ${\tilde {\eta}}$, if the condition
\begin{equation}
E \ge  {m^2 \over \epsilon}
- {3|{\eta}| \over 8} {E^3  \over \epsilon E_{p}}
\label{thr1}
\end{equation}
is satisfied.
This condition defines the actual threshold in the non-systematic-effect
scenario we are considering. Clearly in this sense the threshold
is inevitably decreased by the non-systematic effect.
However, there is only a tiny chance that a given photon
would have ${\tilde {\eta}} = |{\eta}|$, since this is
the limiting case of the range allowed by the uncertainty principle
that is inducing the non-systematic effect, and unless
${\tilde {\eta}} = |{\eta}|$ the process will still not be allowed
even if
\begin{equation}
E \simeq  {m^2 \over \epsilon}
- {3|{\eta}| \over 8} {E^3  \over \epsilon E_{p}} ~.
\label{thr1b}
\end{equation}
Moreover, even assuming ${\tilde {\eta}} = |{\eta}|$,
the energy value described by (\ref{thr1b}) will only be sufficient
to create an electron positron pair with
${\tilde {\eta}}_+ = - |{\eta}|$ and ${\tilde {\eta}}_- = -|{\eta}|$,
which again are isolated points at the extremes of the relevant
probability distributions.
Therefore the process becomes possible at the energy level
described by (\ref{thr1b}) but it remains extremely
unlikely, strongly suppressed by the small probability
that the values of ${\tilde {\eta}}$, ${\tilde {\eta}}_+$
and ${\tilde {\eta}}_-$ would satisfy the kinematical requirements.

With reasoning of this type, one can easily develop
an intuition for the dependence on the energy $E$,
for fixed value of $\epsilon$ (and treating
${\tilde {\eta}}$, ${\tilde {\eta}}_+$
and ${\tilde {\eta}}_-$ as totally unknown), of the likelihood
that the pair-production process can occur:
(i) when (\ref{thr1}) is not satisfied the process is not allowed;
(ii) as the value of $E$ is increased above the
value described by (\ref{thr1b}), pair production becomes
less and less suppressed by the relevant probability
distributions for
${\tilde {\eta}}$, ${\tilde {\eta}}_+$
and ${\tilde {\eta}}_-$,
but some suppression remains up to the value of $E$
that satisfies
\begin{equation}
E \simeq  {m^2 \over \epsilon}
+ {3|{\eta}| \over 8} {E^3  \over \epsilon E_{p}} ~;
\label{thr1c}
\end{equation}
(iii) finally for energies $E$ higher than the one described by
(\ref{thr1c}), the process is kinematically allowed for all
values of ${\tilde {\eta}}$, ${\tilde {\eta}}_+$
and ${\tilde {\eta}}_-$, and therefore the likelihood
of the process is just the same as in the classical-spacetime theory.

Up to this point we have analyzed
a single photon-photon collision, focusing on the kinematic
requirements for electron-positron pair production.
This process is relevant for understanding
certain astrophysical observations, such as the
Mk501 photons. For a Mk501 photon with energy of some 10 or 20 TeV,
there are many opportunities to collide with soft photons
with energy suitable for pair production to occur (the mean
free path is much shorter than the distance between the source
and the Earth). Thus one expects (in the
same sense illustrated through the analogy discussed in the
previous Subsection for the case of many holes between points $A$
and $B$) that even a small probability of producing an electron-positron
pair in a single collision would be sufficient
to lead to the disappearance of the MK501 hard photon before
reaching our detectors. The probability is small in a
single collision with a soft background photon, but
the fact that there are, during the long journey,
many such pair-production opportunities
renders it likely that in one of the many collisions
the hard photon would indeed disappear into an electron-positron
pair. For the specific scheme of non-systematic effects
which we have used as illustrative example in this Subsection,
it appears therefore that a characteristic prediction
is that the Mk501 should start being sharply suppressed
already at the energy level
described by (\ref{thr1b}), below the threshold corresponding to the
classical-spacetime kinematics.
Of course, completely analogous arguments apply to
the analysis of ultra-high-energy cosmic rays.
Since the present interpretation of these observations is that
the particle flux is unsuppressed even beyond
the classical-spacetime prediction, the type of non-systematic
scheme considered here appears not to provide an explanation
of the data. This of course does not exclude altogether the idea
that non-systematic effects might be responsible for those
observations: the non-systematic scheme considered in this Subsection
corresponds, in the analogy of the preceding Subsection,
only to the case in which
non-systematic effects affect the sizes of the many holes
between points $A$ and $B$, while the wheel diameter is fixed;
but one could look for schemes in which other types
of non-systematic effects are present.
If one could replicate the situation which, in our analogy,
involves fixed sizes for the holes but varying diameters for the wheels
affected by non-systematic effects, the phenomenological
implications for Mk501 observations (and cosmic-ray observations)
would be different: one would still expect a suppression of the
flux to start occurring below the classical threshold, albeit only
a small suppression (which could fall within the accuracy
of present measurements), and the flux would then be expected
to extend even beyong the classical threshold, although
with increasingly large suppression.
In this sense, such an alternative non-systematic-effect
scheme would provide a solution of the threshold paradoxes.

%%%%%%%%%%%%%%%%%%%%%%%%%%%%%%%ADD COMMENTS
%%%%%%%%%%%%%%%%%%%%%%%%%%%%%%%ADD COMMENTS
%%%%%%%%%%%%%%%%%%%%%%%%%%%%%%%ADD COMMENTS
%%%%%%%%%%%%%%%%%%%%%%%%%%%%%%%ADD COMMENTS
%%%%%%%%%%%%%%%%%%%%%%%%%%%%%%%ADD COMMENTS
%\subsection{Solving both the Mk501-photon and the UHE cosmic-ray paradoxes
%with non-systematic effects}

%bla bla bla bla bla bla bla bla bla bla bla bla bla bla bla bla
%bla bla bla bla bla bla bla bla bla bla bla bla bla bla bla bla
%bla bla bla bla bla bla bla bla bla bla bla bla
%bla bla bla bla bla bla bla bla bla bla bla bla bla bla bla bla
%bla bla bla bla bla bla bla bla bla bla bla bla

\subsection{Interplay between systematic and non-systematic
effects}

Our analysis of the example of non-systematic effects on particle
kinematics may also be relevant for certain systematic-effect
schemes for solving the threshold paradoxes.
In fact, it appears plausible~\cite{gacdsr} that in quantum spacetime
one would find deformed symmetries, possibly
leading to a systematic
deformation of the dispersion relation,
as well as new uncertainty principles,
possibly affecting in non-systematic fashion the $E \leftrightarrow p$
relation established by the dispersion relation.
It is therefore important to understand the implications
of non-systematic effects, and in particular we should try
to understand how systematic effects and nonsystematic effects
can combine in physical contexts such as the ones
pertaining to the paradoxes we have considered.

As discussed in the analogy introduced in Subsection 3.1,
nonsystematic effects can effectively counter-act
the implications of systematic effects of comparable magnitude.
Following the reasoning described in Subsection 3.1,
the reader can easily realize that it is plausible that
one can have a theory in which spacetime symmetries would
be such that the dispersion relation is deformed in the
direction of inducing an increase of the threshold
with respect to the classical-spacetime prediction,
but then there might also be non-systematic effects
that (with however small probability) can render
particle-production processes possible even below the
new threshold.

A systematic effect may well raise the threshold for
a particle-production process, but in the presence of an
accompanying non-systematic effect, this increase of the threshold
will have to be interpreted only in a statistical sense:
processes with energetics below the new higher threshold can still
(with however small probability) occur.
If the size of the upward shift in the threshold is comparable to
the size of the uncertainties to be taken into account in
threshold analysis, the minimum required energetic configuration
required by the process will be basically unchanged; the only net
implication of the two effects combined would be a probabilistic
modulation (of the type discussed in the previous Subsections)
of the particle-production probability.
When many collision opportunities are
encountered by a high-energy particle on its way
from the source to the Earth, even a small probability
of particle production in each collision may be
sufficient for particles below threshold to be involved
in at least one particle-production process, and this can
counter-act the upward shift of the threshold induced
by the systematic effects.

\section{Closing remarks}

The illustrative example of quantum-gravity-induced threshold
anomalies has allowed us to show that the analysis
of non-systematic quantum-gravity effects can be non-trivial
and neglecting such effects in favour of accompanying systematic
effects may lead to unreliable conclusions.
The stochastic/probabilistic nature of the effects induced
by quantum-gravity uncertainty principles can
be strengthened in some cases by the fact that several
such probabilistic imprints
can characterize a given observation.
For example, in certain models the probability of particle-production
below threshold may be extremely small, but in certain situations,
as in the case of
high-energy particles reaching us
from far away astrophysical sources, this small probability
applies only to a single opportunity of collision with a
soft background photon, while the overall probability of particle production,
considering that a large number of particle-production
opportunities are encountered by the high-energy particle
on its way to our detectors, may still be large.

In light of our analysis, it appears
that attempts to explain the debated threshold anomalies
for cosmic rays and Mk501 photons that rely exclusively on
non-systematic effects are probably problematic; but this type of solution
cannot be excluded in general a priori, as the careful reader
will deduce from some of the remarks in Subsection 3.1.
We have also provided motivations for a more careful study of threshold
anomalies in contexts in which both systematic and non-systematic
effects are expected, especially since
it is not always possible to assume that non-systematic effects
are negligible with respect to systematic ones.

\bigskip

{\bf Acknowledgments}\\

G.A.-C. greatfully acknowledges conversations with Tsvi Piran.
This work was supported in part by the U.S. Department of Energy and
the Bahnson Fund of the University of North Carolina.

\bigskip

%\begin{figure}[p]
%\begin{center}
%\epsfig{file=figlid1.eps,height=14truecm}\quad
%\end{center}
% %\bigskip
%\caption{The region of the $\alpha,\eta$ parameter space that provides a
%solution to both the UHECR and TeV-$\gamma$ threshold anomalies while
%({\it i.e.} the $1/\alpha \rightarrow 0$) limit.}
%\label{fig1}
%\end{figure}

\baselineskip 12pt plus .5pt minus .5pt


\begin{thebibliography}{99}

\bibitem{polonpap} G. Amelino-Camelia, ``Are we at
the dawn of quantum-gravity phenomenology?",
gr-qc/9910089, Lect.~Notes Phys.~{541}, 1-49 (2000).
%notes based on
%lectures given at the XXXV Karpacz Winter School of Theoretical
%Physics {\it From Cosmology to Quantum Gravity}, Polanica, Poland,
%2-12 February, 1999 (published in the volume
%entitled ``Towards Quantum Gravity'', ed: J.~Kowalski-Glikman,
%Springer-Verlag Heidelberg 2000).

\bibitem{iucaapap}
%\bibitem{gacQM100}
G.~Amelino-Camelia, gr-qc/0012049,
Nature 408, 661 (2000);
G.~Amelino-Camelia, ``Quantum-Gravity
Phenomenology: Status and Prospects",
gr-qc/0204051
(invited Brief Review to appear in a
special issue of Modern Physics Letters A devoted to the First
IUCAA Meeting on the Interface of Gravitational and Quantum Realms).

\bibitem{qgcpt}
J.~Ellis, J.S.~Hagelin, D.V.~Nanopoulos and M.~Srednicki,
%Search for violations of quantum mechanics.
Nucl.~Phys.~B241, 381 (1984);
%381--405 (1984).
%\bibitem{huetpesk}
P.~Huet and M.E.~Peskin,
% Violation of CPT and quantum mechanics in the $K_0-\bar{K_0}$ system,
Nucl.~Phys.~B434, 3 (1995);
%3-38.
%\bibitem{kostcpt}
V.A.~Kostelecky and R.~Potting,
%CPT, STRINGS, AND MESON FACTORIES.
Phys.~Rev.~{D51}, 3923 (1995);
%3923-3935,1995
%e-Print Archive: hep-ph/9501341
O.~Bertolami, D.~Colladay, V.A.~Kostelecky and R.~Potting,
%CPT VIOLATION AND BARYOGENESIS.
Phys.Lett.~{B395}, 178 (1997);
%\bibitem{ahlucpt}
D.V.~Ahluwalia,
%RECONCILING SUPERKAMIOKANDE, LSND, AND
%HOME-STAKE NEUTRINO OSCILLATION DATA.
%Mod.Phys.Lett.A13:2249-2264,1998.
%    [HEP-PH 9807267]
Mod.~Phys.~Lett.~A13, 2249 (1998).

\bibitem{grbgac} G.~Amelino-Camelia, J.~Ellis, N.E.~Mavromatos,
D.V.~Nanopoulos and S.~Sarkar,
% {\it Tests of quantum gravity from observations
% of $\gamma$-ray bursts},
astro-ph/9712103,
%Nature {393} (1998) 763-765.
Nature {393}, 763 (1998).

\bibitem{billetal} S.D.~Biller et al.,
Phys.~Rev.~Lett.~{83}, 2108 (1999).

\bibitem{gacgwi} G. Amelino-Camelia,
%{\it Gravity-wave interferometers as quantum-gravity detectors},
gr-qc/9808029,
Nature {398}, 216 (1999):
gr-qc/0104086, Nature {410}, 1065 (2001);
%\bibitem{bignapap} G. Amelino-Camelia,
%{\it Gravity-wave interferometers as probes of a
%low-energy effective quantum gravity},
gr-qc/9903080, Phys.~Rev.~{D62}, 024015 (2000).

\bibitem{nggwi} Y.J.~Ng and H.~van Dam,
%MEASURING THE FOAMINESS OF SPACE-TIME WITH GRAVITY-WAVE INTERFEROMETERS
gr-qc/9906003,
Y.J. Ng and H. van Dam, Found.~Phys.~30, 795 (2000);
gr-qc/9911054, Phys.~Lett.~B477, 429 (2000).

\bibitem{gacgwiLATEST} G.~Amelino-Camelia,
%PHENOMENOLOGICAL DESCRIPTION OF SPACE-TIME FOAM.
gr-qc/0104005.

\bibitem{kifu} T.~Kifune,
%{\it Invariance violation extends the cosmic ray horizon?},
%astro-ph/9904164,
Astrophys.~J.~Lett.~{518}, L21 (1999);
%\bibitem{kluz}
W.~Kluzniak,
%{\it Is the Universe transparent to TeV photons?},
%astro-ph/9905308.
%Talk given at 8th International Workshop on Neutrino
%Telescopes, Venice, Italy, 23-26 Feb 1999.
%In *Venice 1999, Neutrino telescopes, vol. 2* 163-166.
%
%TRANSPARENCY OF THE UNIVERSE TO TEV PHOTONS IN SOME MODELS OF
%QUANTUM GRAVITY.
%By W. Kluzniak
%(Wisconsin U., Madison & Warsaw, Copernicus Astron. Ctr.). 1999.
%Prepared for 1st VERITAS Workshop
%on TeV Astrophysics of Extragalactic Sources, Cambridge, MA,
%23-24 Oct 1998.
%Astropart.~Phys.~11 (1999) 117-118.
Astropart.~Phys.~11, 117 (1999);
%\bibitem{aus}
R.J.~Protheroe and H.~Meyer,
%{\it An infrared background TeV gamma ray crisis?},
%astro-ph/0005349.
Phys.~Lett.~{B493}, 1 (2000).
%Phys.~Lett.~B493 (2000) 1-6.

\bibitem{gactp}  G. Amelino-Camelia and T. Piran,
Phys.~Rev.~{D64}, 036005 (2001).

\bibitem{ngthresh}
Y.J.~Ng, D.S.~Lee, M.C.~Oh, and H.~van Dam, Phys.~Lett.~B507, 236 (2001).

\bibitem{gacpion} G. Amelino-Camelia, gr-qc/0107086,
Phys.~Lett.~{B528}, 181 (2002).
%{\it Phys.~Lett.}~B {\bf 528}, 181-187 (2002).

\bibitem{veneziano}
R.~Brustein, gr-qc/9810063;
G.~Veneziano, hep-th/9902097;
M.~Gasperini,
hep-th/9907067;
%\bibitem{perci}
I.C.~Percival and W.T.~Strunz,
Proc.~R.~Soc.~{A453}, 431 (1997);
%\bibitem{garaytest}
L.J.~Garay,
Phys.~Rev.~Lett.~{80}, 2508 (1998);
%\bibitem{ahlunature}
D.V.~Ahluwalia,
%{\it Quantum gravity: testing time for theories},
gr-qc/9903074, Nature {398}, 199 (1999);
%\bibitem{lamer}
C.~L\"{a}mmerzahl and C.J.~Bord\'{e},
%Testing the Dirac equation
%gr-qc/0101095
%to appear in: , C.W.F. Everitt, and F.W. Hehl (eds.): Gyros, Clocks,
%Interferometers...: Testing Relativistic Gravity in Space,
%Lecture Notes in Physics 562, Springer 2001
%Lect.~Notes Phys.~562 (2001) 463-478
{Lect.~Notes Phys.}~{562}, 463 (2001).

\bibitem{rovhisto} C.~Rovelli,
%NOTES FOR A BRIEF HISTORY OF QUANTUM GRAVITY.
gr-qc/0006061 (in Procedings of the 9th Marcel Grossmann Meeting
on Recent Developments in Theoretical and Experimental General Relativity,
Gravitation and Relativistic Field Theories, Rome, Italy, 2-9 Jul 2000).
%\bibitem{carlip}
S.~Carlip,
%QUANTUM GRAVITY: A PROGRESS REPORT.
gr-qc/0108040,
Rept.~Prog.~Phys.~{64}, 885 (2001).

\bibitem{gacdsr} G.~Amelino-Camelia, gr-qc/0012051,
%{\it Relativity in space-times with short-distance
%structure governed by an observer-independent (Planckian)
%length scale},
%Int.~J.~Mod.~Phys.~{\bf D11}, 35 (2002) 35-60;
Int.~J.~Mod.~Phys.~{D11}, 35 (2002);
hep-th/0012238,
%{\it Testable scenario for Relativity with minimum length},
%Phys.~Lett.~{\bf B510} (2001) 255-263.
Phys.~Lett.~{B510}, 255 (2001);
%\bibitem{dsr3}
G.~Amelino-Camelia,
%STATUS OF RELATIVITY WITH OBSERVER INDEPENDENT LENGTH AND VELOCITY SCALES
gr-qc/0106004 (in ``Karpacz 2001, New developments in fundamental
interaction theories", pages 137-150).

\bibitem{GZK} K.~Greisen, Phys.~Rev.~Lett.~16, 748 (1966);
G.~T.~Zatsepin and V.~A. Kuzmin, Sov.~Phys.-JETP Lett.~4, 78 (1966).

\bibitem{NGS} A.I.~Nikishov, Sov.~Phys.~JETP 14, 393 (1962);
J.~Gould, G.~Schreder, Phys.~Rev.~155, 1404 (1967);
F.W.~Stecker, O.C.~De Jager and M.H.~Salomon,
Ap.J.~390, L49 (1992).

\bibitem{AgaWat} M.~Takeda et al., Phys. Rev. Lett.~81, 1163 (1998);
A.~Watson, Proc. Snowmass Workshop, p 126, 1996.

\bibitem{Aharonian99} F.A.~Aharonian et al., A\&A 349, 11A (1999).

\bibitem{Wright00} E.L.~Wright, astro-ph/0004192.

\bibitem{Finkbeiner} D.P.~Finkbeiner, M.~Davis and D.J.~Schlegel,
astro-ph/0004175.

\bibitem{Hauser98} M.G.Hauser, R.G.~Arendt, T.~Kelsall et~al.,
{ApJ} {508}, 25 (1998).

\bibitem{Biviano99} A. Biviano et al., astro-ph/9910314.

\bibitem{Olinto} A.V.~Olinto, Phys.~Rept.~333, 329 (2000).
%A.V.~Olinto, Phys.~Rept.~333-334 (2000) 329-348

\bibitem{steckernoparadox} O.C.~De Jager and F.W.~Stecker,
astro-ph/0107103,
%Extragalactic Gamma-ray Absorption and the
%Intrinsic Spectrum of Mkn 501 During the 1997 Flare
%Astrophys.J. 566 (2002) 738-743
Astrophys.J. 566 (2002) 738.

\bibitem{Gonzalez} L. Gonzalez-Mestres, physics/9704017;
%\bibitem{colgla}
S.~Coleman, S.L.~Glashow,
%{\it High-energy tests of Lorentz invariance},
Phys.~Rev.~D59, 116008 (1999).

\bibitem{padma} T.~Padmanabhan,
Class.~Quantum Grav.~4, L107 (1987);
%\bibitem{dvaUP}
D.V.~Ahluwalia,
%{\it Quantum measurement, gravitation, and locality},
%Phys. Lett. {\bf B339} (1994) 301-303.
Phys.~Lett.~B339, 301 (1994);
%\bibitem{ngmpla}
Y.J.~Ng and H.~van Dam,
Mod.~Phys.~Lett~A9, 335 (1994);
%\bibitem{gacmpla}
G.~Amelino-Camelia,
Mod.~Phys.~Lett.~A9, 3415 (1994);
Y.J.~Ng and H.~van Dam, Mod.~Phys.~Lett.~A10, 2801 (1995);
G.~Amelino-Camelia, Mod.~Phys.~Lett.~A11, 1411 (1996).
%Lett. {\bf A9} (1994) 3415--3422; {\it ibid.} {\bf A11} (1996) 1411--1416.
%{\it Limits on the measurability of space-time distances in the
%  semi-classical approximation of quantum gravity},
%Mod. Phys. Lett. {\bf A9} (1994) 3415-3422.

\end{thebibliography}
\end{document}